\documentclass[preprint,showpacs,preprintnumbers,amsmath,amssymb]{revtex4}

\usepackage{graphicx}
\usepackage{dcolumn}
\usepackage{bm}
\renewcommand{\Re} {{\rm Re}}

\begin{document}
\draft


\title{Giant transmission and dissipation in perforated films mediated by surface
phonon polaritons}

\author{D.~Korobkin}
\author{Y.~Urzhumov}
\author{B.~Neuner III}
\author{G.~Shvets}
\email{gena@physics.utexas.edu}
\affiliation{Department of Physics, The University of Texas at
Austin, Austin, Texas 78712 }

\author{Z.~Zhang and I.~D.~Mayergoyz}
\affiliation{Department of Electrical and Computer Engineering,
University of Maryland, College Park, MD 20742}

\date{\today}

\begin{abstract}
We experimentally and theoretically study electromagnetic
properties of optically thin silicon carbide (SiC) membranes
perforated by an array of sub-wavelength holes. Giant absorption
and transmission is found using Fourier Transformed Infrared
(FTIR) microscopy and explained by introducing a
frequency-dependent effective permittivity $\epsilon_{\rm
eff}(\omega)$ of the perforated film. The value of $\epsilon_{\rm
eff}(\omega)$ is determined by the excitation of two distinct
types of hole resonances: a delocalized slow surface polariton
(SSP) whose frequency is largely determined by the array period,
and a localized surface polariton (LSP) which corresponds to the
resonances of an isolated hole. Only SSPs are shown to modify
$\epsilon_{\rm eff}(\omega)$ strongly enough to cause giant
transmission and absorption.
\end{abstract}

\maketitle

Diffraction of light is the major obstacle to increasing the
density of optical circuits and integrating them with
electronics~\cite{ozbay_science06}: light cannot be confined to
dimensions much smaller than half of its wavelength $\lambda/2$.
Utilizing materials with a negative dielectric permittivity
circumvents diffraction limit because interfaces between
polaritonic ($\epsilon < 0$) and dielectric ($\epsilon > 0$)
materials support quasi-electrostatic waves (surface polaritons)
that can be confined to sub-$\lambda$ dimensions. Negative
$\epsilon$ can be due to either collective oscillation of
conduction electron in metals (plasmons)~\cite{barnes_nature03} or
lattice vibrations in polar crystals such as ZnSe, SiC,
InP~\cite{hillenbrand_nature04}. Both metallic and SiC interfaces
have been shown to guide surface plasmon (and phonon) polaritons
over long distances~\cite{berini_prb01,zia_josa04,huber_apl05},
support negative-index
waves~\cite{shvets_prb03,shih_prl06,alu_josab06}, and serve as
elements of a
super-lens~\cite{zhang_science05,melville_optexp05,korobkin_josab06}.
Therefore, the explosion of research aims at understanding basic
polaritonic components: grooves in metal
films~\cite{bozhevolniy_nature06}, nanoparticle-film
polaritons~\cite{le_nano05}, hole arrays~\cite{ebbesen98}, and
single holes~\cite{kall_nanolett04}.

Sub-$\lambda$ hole arrays were the first to attract significant
attention after the discovery of extraordinary optical
transmission~\cite{ebbesen98} through perforated optically thick
metallic films and the ensuing
controversy~\cite{lalanne_prl02,lezec_optexp04} about the role of
surface polaritons in transmission enhancement. The difficulty
with interpreting the role of surface polaritons (SPs) in
transmission through optically thick ($\lambda_{\rm skin} \equiv
\lambda/4\pi\sqrt{|\epsilon|} \ll H$, where $H$ is the film
thickness) films arises from the fact that even holes in a perfect
conductor (which does not support surface polaritons) can spoof
surface polaritons~\cite{pendry_science04}, especially when
$\lambda$ is close to array period $L$. No such spoofing occurs
for optically thin films.

In this Letter we report results of the spectroscopic study of
such suspended perforated SiC membranes, with both hole diameter
$D \ll \lambda$ and period $L < \lambda$ significantly
sub-$\lambda$. SiC is chosen because of its low
losses~\cite{hillenbrand_nature04} and the possibility of growing
high quality thin SiC films on Si substrates~\cite{zorman95}.
While the majority of surface phonon polariton experiments to-date
have been conducted with SiC
substrates~\cite{greffet_nature02,hillenbrand_nature04,huber_apl05},
thin SiC membranes hold a greater promise for fabricating novel
multi-layer nanophononic structures~\cite{korobkin_josab06}. Using
FTIR microscopy, we demonstrate that, in adjacent frequency
ranges, giant transmission and absorption of the incident
radiation is realized. Both phenomena are shown to be due to the
excitation of quasi-electrostatic SPs. It is shown that a
perforated film can be described as a metamaterial with the
effective  permittivity $\epsilon_{\rm eff} = \epsilon_r +
i\epsilon_i$ strongly modified by the excitation of SPs: regions
of giant transmission and absorption are related, respectively, to
the lowering of $|\epsilon_r|$ and increase of $\epsilon_i$. In
addition, we show theoretically that two types of SPs are
supported by the hole arrays in a polaritonic membrane: localized
surface polaritons (LSPs) and delocalized SSPs.

\begin{figure}
\includegraphics[width=85mm]{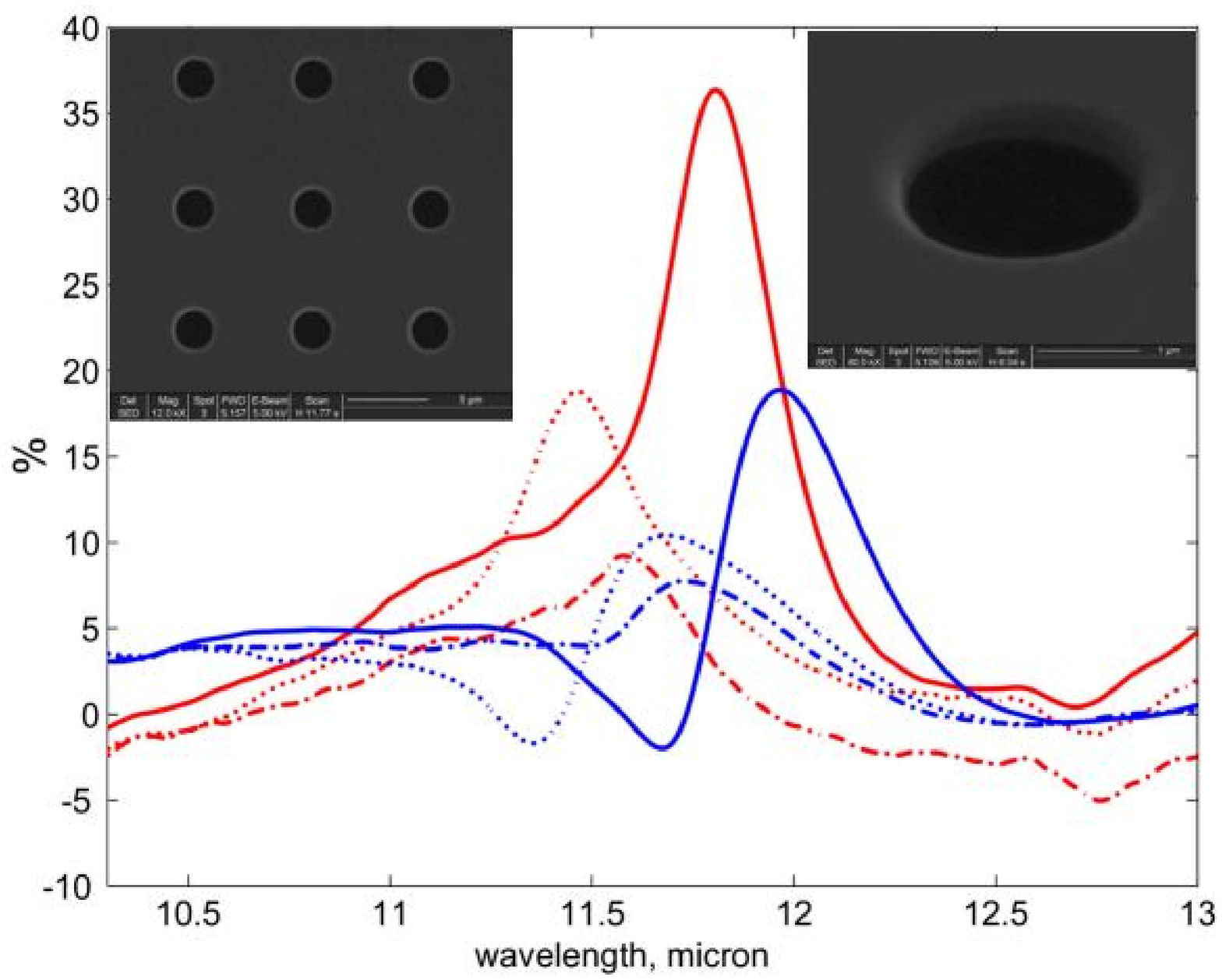}
 \caption{\label{fig:trans_experiment} (Color online) Extra
 transmission (blue) and absorption (red) of square arrays of
 circular holes in a $458$nm thick SiC film, relative to
 non-perforated film. Solid: $L=7\mu$m, $D=2\mu$m; dash-dotted:
 $L=7\mu$m, $D=1\mu$m; dotted: $L=5\mu$m, $D=1\mu$m. Left inset: SEM
 image of the $L=7\mu$m, $D=2\mu$m sample. Right inset: SEM image
 of a $D=2\mu$m hole (at $52^\circ$ from normal). }
\end{figure}

Square arrays of round holes were milled by a dual focused ion
beam (FIB) system (FEI Strata 235) in a suspended SiC membrane.
The starting material for membrane fabrication was a $458$ nm
thick single-crystalline 3C-SiC film heteroepitaxially grown on
the front (polished) side of a $0.5$ mm thick Si$(100)$
wafer~\cite{zorman95}. The $0.35 \times 0.35$ mm suspended
membrane was produced by anisotropic KOH etching of the Si from
the back (unpolished) side~\cite{korobkin_josab06}. All holes were
round in shape (see insets to Fig.~\ref{fig:trans_experiment}),
with the edge radius of curvature of order $30$ nm. The hole
diameter and period in different samples varied in the range $1
\le D \le 2 \mu$m and $5 \le L \le 7 \mu$m, and a typical
perforated area was $150\times150 \mu$m in size. Inset of the
Fig.~\ref{fig:trans_experiment} shows one fabricated sample with
$D=2 \mu$m and $L=7 \mu$m. An FTIR microscope (Perkins-Elmer
Spectrum GX AutoImage) with spatial domain $(100 \mu$m$)^2$ has
been used to measure the transmission and reflection from the
perforated and non-perforated areas in the $1500-700$cm$^{-1}$
spectral range. The FTIR microscope utilized Cassegrain optics
with the angle of incidence between $9^\circ-35^\circ$. Input IR
radiation was polarized by a wire-mesh (ISP Optics) polarizer.
To quantify the effect of the holes, transmission and absorption
spectra of a non-perforated SiC membrane are subtracted from the
corresponding hole array spectra and plotted in
Fig.~\ref{fig:trans_experiment} for several values of $D$ and $L$.

Below we focus on the ($D=2\mu$m, $L=7\mu$m) sample. Although the
deeply sub-$\lambda$ holes occupy only $\eta = 6\%$ of the total
sample area, they increase the transmittance through the sample by
$T_{\rm h} = 20\%$ (from $T_{\rm film} = 10\%$ to $T_{\rm perf} =
30\%$) at $\lambda_{tr} = 11.95\mu$m. This $\times3$ transmission
increase over the fractional holes' area is quite remarkable;
outside of the {\it reststrahlen} ($\epsilon(\omega) < 0$) region
of SiC the increase is much smaller than $\eta$. Even more
remarkable is the increase of absorption (up by almost $40\%$ from
$< 1\%$ in a non-perforated film) at $\lambda_{abs} = 11.8\mu$m.
Other perforated samples ($L=5\mu$m, $D=1\mu$m) and ($L=7\mu$m,
$D=1\mu$m) were also experimentally studied and gave qualitatively
similar results as shown in Fig.~\ref{fig:trans_experiment}. The
absorption maximum is blue-shifted for smaller holes
($\lambda_{abs} = 11.57\mu$m for the $L=7\mu$m, $D=1\mu$m sample)
and for smaller periods ($\lambda_{abs} = 11.49\mu$m for the
$L=5\mu$m, $D=1\mu$m sample), in accordance with the theory
presented below which explains the anomalies in transmission and
absorption as manifestations of surface phonon polariton
excitation. This theory paves the way to the engineering of
optical properties of polaritonic films using resonant excitation
of surface polaritons. More complex three-dimensional materials
can then be built using multi-layer polaritonic films as building
blocks.

Because of the sub-$\lambda$ nature of the perforated film, it can
be described as an {\it effective medium} with a
frequency-dependent permittivity $\epsilon_{\rm eff}(\omega)$
determined by several strongest electrostatic (ES) resonances.
Strengths and frequencies of these resonances are numerically
computed using two different formalisms: generalized eigenvalue
differential equation
(GEDE)~\cite{bergman92,stockman_bergman01,shvets04} and the
surface integral eigenvalue equation~\cite{fredkin03,mayergoyz05}.
The former method has been successfully applied to periodic
sub-$\lambda$ polaritonic crystals consisting of non-connected
polaritonic inclusions~\cite{shvets04}. To the best of our
knowledge, this is the first generalization of this method to the
system with a {\it continuous} polaritonic phase.

\begin{figure}
\includegraphics[width=85mm]{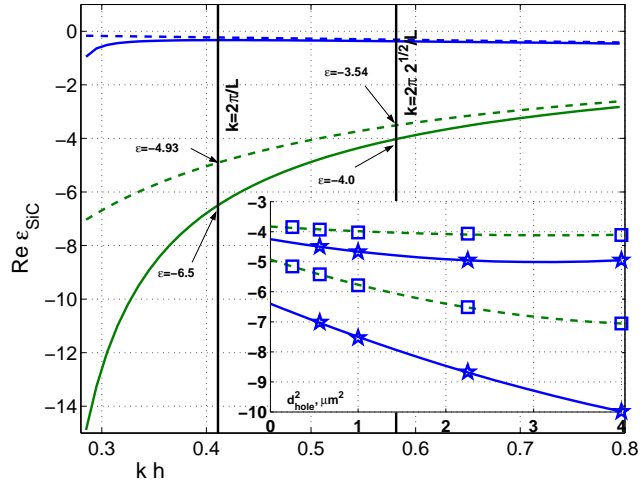}
 \caption{\label{fig:dispersion} (Color online) Dispersion relation of the even
 (green, lower curves) and odd (blue, upper curves) surface
 polaritons on a SiC film with thickness $H = 458$ nm.
 Solid lines: exact dispersion relation in the form of $\Re [\epsilon_{\rm
 SiC}(\omega)]$~vs.~$k$, dashed lines: ES approximation. Two
 vertical lines: $k=2\pi/L$ and $k=2\sqrt{2}\pi/L$ for $L=7 \mu$m.
 Inset: position of delocalized resonances SSP$(1,0)$ and SSP$(1,1)$
 as a function of hole diameter $D$ (squares on dashed line --
 ES eigenvalue simulations; stars on solid line -- absorption peak
 positions in EM FEFD simulations).}
\end{figure}

The details of the GEDE approach as applied to periodic
nanostructures consisting of two material components (one with a
frequency-dependent $\epsilon(\omega) < 0$ and another with
$\epsilon_d = 1$) is described
elsewhere~\cite{bergman92,shvets_jopa05}. Briefly, the following
steps are followed. First, the GEDE $\vec{\nabla} \cdot \left[
\theta(\vec{x}) \vec{\nabla} \phi_i \right] = s_i \nabla^2 \phi_i$
is solved for the real eigenvalue $s_i$, where $\theta(\vec{x}) =
1$ inside the polaritonic material and $\theta(\vec{x}) = 0$
elsewhere, and $\phi_i$ is a potential eigenfunction with the
dipole symmetry periodic with period $L$ in the plane of the film.
Second, the dipole strengths $f_i$ proportional to the squared
dipole moment of each resonance are
calculated~\cite{stockman_bergman01}. Finally, the ES permittivity
$\epsilon_{\rm eff}(\omega)$ is found by summing up the
contributions of all dipole-active resonances:
\begin{equation}\label{eq:eps_eff}
    \epsilon_{\rm eff}(\omega) = 1 - \frac{f_0}{s(\omega)} -
    \sum_{i>0} \frac{f_i}{s(\omega)-s_i},
\end{equation}
where $s = [1-\epsilon(\omega)]^{-1}$ serves as a frequency label.
Note the pole at $s=0$ (corresponding to $\epsilon\to-\infty$) in
Eq.~(\ref{eq:eps_eff}) which emerges due to {\it continuous}
polaritonic phase. ``Frequencies'' $s_i$ and strengths $f_i$ of
the dominant dipole resonances computed for the experimentally
relevant parameters ($D=2\mu$m, $H=458$nm, and $L=7 \mu$m) are:
$s_0=0$ ($\epsilon_0=-\infty$) with $f_0\approx 0.88$, $s_1\approx
0.1241$ ($\epsilon_1\approx-7.058$) with $f_1\approx 0.041$,
$s_2\approx 0.1958$ ($\epsilon_2\approx-4.107$) with $f_2\approx
0.0054$, $s_3\approx 0.255$ ($\epsilon_3\approx -2.922$) with
$f_3\approx 0.0036$, and $s_4\approx 0.666$
($\epsilon_4\approx-0.50$) with $f_4\approx 0.0049$. Other hole
diameters were also studied, and the corresponding
$\epsilon_{1,2}$ are indicated by squares in the inset of
Fig.~\ref{fig:dispersion}. Note that $\epsilon_{\rm eff}(\omega)$
is always finite because $s$ is a complex number due to finite
losses in SiC while all $s_i$ are real.

\begin{figure}
\includegraphics[width=85mm]{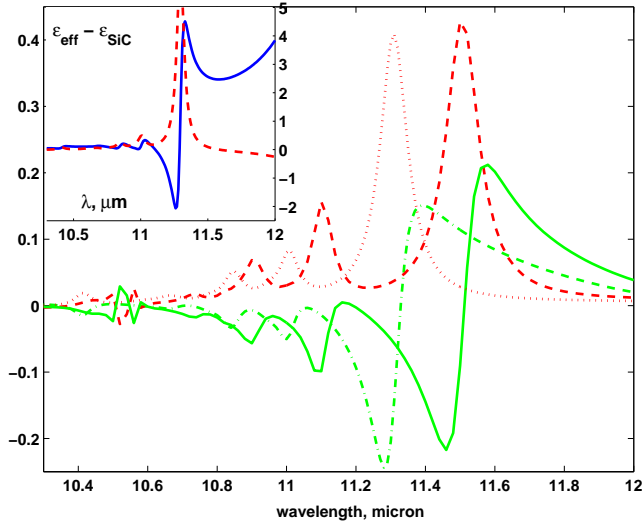}
 \caption{\label{fig:transmission} (Color online) Theoretical extra transmission
 (green solid and dash-dotted curves) and absorption (red dashed and dotted curves)
 of a $7\mu$m array of $2 \mu$m holes in $458$ nm film of SiC.
 Solid and dashed curves: FEFD simulation of EM wave scattering;
 dash-dotted and dotted: theoretic estimate based on
 ES $\epsilon_{\rm eff}$, which is plotted on the inset.}
\end{figure}

Experimental results depicted in Fig.~\ref{fig:trans_experiment}
can now be interpreted using $\epsilon_{\rm eff}(\omega)$: the
normal incidence transmission and reflection coefficients through
a slab of thickness $H$ are given by
\begin{equation}\label{eq:transmission}
    T = \left| \frac{(1-r_1^2)e^{ik_0(n-1)H}}{1 - r_1^2 e^{2ik_0
    n H}} \right|^2, \ \ R = \left| \frac{r_1(1-e^{2ik_0nH})}{1 - r_1^2 e^{2ik_0
    n H}} \right|^2,\nonumber
\end{equation}
where $n = \sqrt{\epsilon_{\rm eff}}$, $k_0=\omega/c$ and $r_1 =
(1-n)/(1+n)$. The standard polaritonic formula recently
verified~\cite{korobkin_josab06} for suspended SiC membranes was
used for $\epsilon_{\rm SiC}$. We have calculated transmittance
$T$ and absorbance $A=1-R-T$ of the perforated film, subtracted
the corresponding quantities for the non-perforated film
($\epsilon_{\rm eff}$ replaced by $\epsilon_{\rm SiC}$), and
plotted the corresponding differential quantities in
Fig.~\ref{fig:transmission}. To facilitate the interpretation of
the giant transmission and absorption effects, a segment of ${\rm
Re}[\epsilon_{\rm eff}],{\rm Im}[\epsilon_{\rm eff}]$ dependence
is plotted in the inset to Fig.~\ref{fig:transmission} near the
strongest resonance at $\lambda_1 = 11.3 \mu$m, where ${\rm
Re}[\epsilon_{\rm SiC}(\lambda_1)] = \epsilon_1$.

The absorption spike (dotted line) is due to the peak of ${\rm
Im}[\epsilon_{\rm eff}]$ at $\lambda = \lambda_1$. The
transmission maximum occurs due to the decrease of the absolute
value of ${\rm Re}[\epsilon_{\rm eff}]$ at $\lambda =
\lambda_{max}$ shown in the inset (solid line) to
Fig.~\ref{fig:transmission}. Note that the enhanced transmission
predicted from $\epsilon_{\rm eff}$ occurs for $\lambda_{max} >
\lambda_1$ in agreement with experimental observations. Other much
smaller absorption and transmission peaks are also predicted.
However, experimental accuracy was not sufficient to identify
these additional variations of $T$ and $A$ in
Fig.~\ref{fig:trans_experiment}. Because spacing between the holes
($L = 7 \mu$m) was not negligible compared with the wavelength
($\lambda \sim 11\mu$m), we have carried fully electromagnetic
(EM) calculations of $T$ and $A$ using the finite-elements
frequency domain (FEFD) solver COMSOL. The results displayed in
Fig.~\ref{fig:transmission} show qualitative agreement with the
$\epsilon_{\rm eff}$-based calculation, with the exception that
all transmission and absorption maxima are slightly red-shifted.
This red shift is a previously noted~\cite{mayergoyz05} phenomenon
explained by the EM corrections to the purely ES response of
sub-$\lambda$ polaritonic structures.

\begin{figure}
\includegraphics[width=85mm]{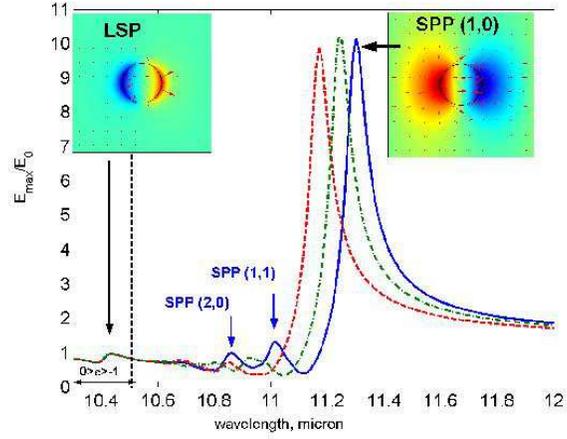}
 \caption{\label{fig:resonances} (Color online) Electric field
 enhancement in the symmetry plane of a SiC film perforated
 with a $L=7\mu$m-period square array of $D=2 \mu$m
 round holes. Insets: ES potential profile at the
 resonances: (left) LSP resonance, and (right) SSP$(1,0)$ resonance.}
\end{figure}

One manifestation of resonances of $\epsilon_{\rm eff}(\omega)$ is
an increase of the peak electric field $E_{\rm max}$ inside the
hole in response to the applied across the structure ac electric
field with amplitude $E_0$. The ratio of $E_{\rm max}/E_0$ is
plotted in Fig.~\ref{fig:resonances} as a function of
 $\lambda$ for a fixed hole diameter $D = 2\mu$m and
variable period $L$. Four enhancement spikes corresponding to
resonances of $\epsilon_{\rm eff}(\omega)$ can be identified for
all periods. Frequencies of the three red resonances ($\lambda
> 10.5 \mu$m) are located in the ${\rm Re}[\epsilon_{\rm SiC}] < -1$ band and are all strongly dependent on $L$.
On the contrary, blue
resonance ($\lambda_{\rm loc} \approx 10.45 \mu$m) belongs to the
$-1 < {\rm Re}[\epsilon_{\rm SiC}] < 0$ range and is
period-independent.
Another striking difference between the red and blue resonances is
that the former  are very de-localized, while the latter  is
strongly localized near the hole (see the two insets to
Fig.~\ref{fig:resonances}).
Identification of the period-independent resonance as the LSP of a
single hole, and of the period-dependent resonances as
de\-localized waves related to SSPs of the smooth film is one of
the main theoretical advances of this Letter.

To validate this identification, we start with reviewing the
properties of SPs of a smooth thin negative-$\epsilon$ film
surrounded by vacuum. Through symmetry, SPs can be labeled by the
parity of the in-plane electric field with respect to the
mid-plane, and by a continuous in-plane wavenumber $k$. A
dispersion relation $\omega$ vs.~$k$ in the ES limit of $k \gg
\omega/c$ is given implicitly by $\epsilon_{\pm}(\omega)=-\tanh(k
H/2)^{\mp 1}$, where $\pm$ refers to even and odd modes,
respectively. Note that even (``slow'') modes are located in the
$-\infty<\epsilon<-1$ part of the spectrum, while the odd
(``fast'') modes are in the $-1<\epsilon<0$ range. EM corrections
to the dispersion relation do not alter this ordering (except for
$k \approx \omega/c$ case that is not relevant to this work). The
EM dispersion curve and its ES approximation (expressed as ${\rm
Re}[\epsilon_{\rm SiC}](\omega)$ vs.~$k$) for the fast and slow
SPs on a smooth SiC film are plotted in Fig.~\ref{fig:dispersion}.

Because $k > \omega/c$ for SPs, they cannot be excited by an EM
wave incident from vacuum. A {\it perforated} film, however, acts
as a crossed diffraction grating and provides coupling between
normally incident radiation (with almost uniform electric field
$\vec{E}_0 = E_0 \vec{e}_x$) and surface modes of the film with
the in-plane wavenumbers $k_{(m,n)}=\left|\frac{2\pi}{L}(m \hat
x+n \hat y)\right|
\equiv\frac{2\pi}{L}\sqrt{m^2+n^2}$~\cite{ebbesen04}. Resonances
of $\epsilon_{\rm eff}(\omega)$ can be understood as the result of
 strong coupling of $\vec{E}_0$ to {\it slow} surface polaritons
(SSP$(m,n)$~\cite{ebbesen04}). This interpretation is verified by
extrapolating the diameter-dependent resonant $\epsilon$ for the
two low-frequency eigenmodes to $D=0$ shown in the inset to
Fig.~\ref{fig:dispersion}. It is apparent from
Fig.~\ref{fig:dispersion} that these two resonances correspond to
excitation of SSP($0,1$) and SSP($1,1$) of the smooth film.
Another conclusion that can be drawn from
Fig.~\ref{fig:dispersion} is that EM effects shift the frequencies
of SPs towards red. This explains the red shift of transmission
and absorption spikes in EM simulations from their positions in ES
model.

No identification with one of the smooth film SSPs can be made
for the LSP resonance at $\lambda = 10.45\mu$m. In fact, this
highly localized near the hole perimeter surface wave resonance
can be thought of as an even-parity ``defect state'' created by
the presence of a single hole in a negative-$\epsilon$ film.
Because the frequency range for which $-1 < \epsilon(\omega) < 0$
is a stop-band for even-parity propagating SPs, the even-parity
LSP can exist in this, and only this frequency range. Because of
the localized nature of the LSP, its frequency is insensitive to
the proximity of other holes (i.~e.~to the period $L$) but is
sensitive to the aspect ratio of the hole and even to the radius
of curvature of hole edge.

In conclusion, we have experimentally demonstrated giant
absorption and transmission through optically thin SiC membranes
perforated by a rectangular array of round holes. These effects
are theoretically explained by introducing the effective
 permittivity of a perforated membrane with a complex
frequency dependence due to resonant coupling to almost-ES surface
phonon polaritons. Two types of phonon polariton resonances are
theoretically uncovered: (i) delocalized modes related to the SSPs
of a smooth film, and (ii) a localized surface polariton (LSP) of
a single hole that exists in the spectral range complementary to
that of SSPs.

This work is supported by the ARO MURI W911NF-04-01-0203 and the
DARPA contract HR0011-05-C-0068. We gratefully acknowledge
Drs.~A.~Aliev and A.~A.~Zakhidov for their assistance with FTIR
microscopy and Dr.~C.~Zorman for growing SiC films and useful
discussions.


\newpage

\newpage

\newpage

\newpage

\end{document}